\documentstyle[12pt,prc,aps,epsfig]{revtex}
\tighten
\begin{document}
\draft
\title{The $\omega\sigma\gamma$-vertex in light cone QCD sum rules}

\author{A. Gokalp~\footnote{Electronic address: agokalp@metu.edu.tr} and
        O. Yilmaz~\footnote{Electronic address: oyilmaz@metu.edu.tr}}
\address{ Physics Department, Middle East Technical University,
         06531 Ankara, Turkey}
\date{\today}
\maketitle

\begin{abstract}
We investigate  the $\omega\sigma\gamma$-vertex and estimate the
coupling constant g$_{\omega\sigma\gamma}$ in the framework of the
light cone QCD sum rules. We compare our result with the values of
this coupling constant deduced from a phenomenological analysis of
$\omega\rightarrow\pi\pi\gamma$ decays.
\end{abstract}

\pacs{PACS numbers: 12.38.Lg;13.40.Hq;14.40.Aq }

The determination of the hadronic observables from experimental
data, such as coupling constants and form factors, requires
physics information  at large distances. However, such information
cannot be obtained directly from the fundamental QCD Lagrangian in
this nonperturbative domain. Among the various nonperturbative
methods, traditional QCD sum rules \cite{R1} have proved very
useful in studying the properties of low-lying hadrons. Further
progress has been achieved by an alternative method known as the
QCD sum rules on the light cone \cite{R2,R3,R4,R5} based on the
operator product expansion on the light cone, which is an
expansion over the twist of the operators rather than dimensions
as in the traditional QCD sum rules. In the present work, we study
the ${\omega\sigma\gamma}$-vertex and we estimate the coupling
constant g$_{\omega\sigma\gamma}$. The traditional QCD sum rules
approach requires for its applicability that in all the channels
of the process under consideration the virtuality should be large
\cite{R1}. However, for the ${\omega\sigma\gamma}$-vertex, the
virtuality in gamma channel is not large. For this reason, in the
present calculation we employ the light cone QCD sum rules method.

The low-mass scalar mesons have fundamental importance in the
phenomenology of low energy QCD and from the point of view of
hadron spectroscopy. Over the years, experimental evidence has
accumulated for their existence \cite{R6,R7} and different
proposals about their nature and  about their quark substructure
have been put forward. At present, whether they are conventional
$q\overline{q}$ states \cite{R8}, $\pi\overline{\pi}$,
$K\overline{K}$ molecules \cite{R9}, or multiquark
$q^2\overline{q}^2$ states \cite{R10} is still a subject of
debate. On the other hand, they are relevant hadronic degrees of
freedom, and therefore the role they play in hadronic processes
should also be studied besides the questions of their nature.

The $\omega\sigma\gamma$-vertex has importance in different areas
of hadron physics. In the calculation of the electromagnetic form
factors of the deuteron,  $\omega\sigma\gamma$-vertex plays a
special role \cite{R11}. The $\omega\sigma\gamma$-exchange current
compensates the large effect of the $\rho\sigma\gamma$-exchange
current contribution which is commonly included in calculations of
elastic electron-deuteron scattering, and therefore the knowledge
of the coupling constant g$_{\omega\sigma\gamma}$ is essential in
such studies. Furthermore, at low energies near threshold in the
electromagnetic production reactions of vector mesons on nucleon
targets, scalar and pseudoscalar meson exchange mechanisms become
important \cite{R12}, and in particular the coupling constant
g$_{\omega\sigma\gamma}$  may be needed as a physical input for
the studies of photoproduction of $\omega$-mesons on nucleons near
threshold.

In order to study the $\omega\sigma\gamma$-vertex and to estimate
the coupling constant g$_{\omega\sigma\gamma}$, we consider the
two point correlation function with photon
\begin{equation}\label{e1}
  T_{\mu}(p,q)=i\int d^{4}x e^{ip\cdot x}
  <\gamma(q)|T\{j_\mu^\omega(x)j_{\sigma}(0)\}|0>
\end{equation}
where
$j_{\mu}^\omega=\frac{1}{6}(\overline{u}^a\gamma_{\mu}u^a+\overline{d}^a\gamma_{\mu}d^a)$
and $j_{\sigma}=\frac{1}{2}(\overline{u}^bu^b+\overline{d}^bd^b)$
are the interpolating currents for for $\omega$ and $\sigma$
mesons with u and d denoting up and down quark fields,
respectively, and a, b are the color indices. The overlap
amplitudes of these interpolating currents with the meson states
are defined as
\begin{eqnarray}\label{e2}
<0|j_{\mu}^\omega|\omega>&=&\lambda_\omega u_\mu \nonumber \\
<0|j_{\sigma}|\sigma>&=&\lambda_\sigma
\end{eqnarray}
where $u_\mu$ is the polarization vector of $\omega$ meson. The
coupling constant  g$_{\omega\sigma\gamma}$ is defined through the
effective Lagrangian
\begin{equation}\label{e3}
{\cal L}=\frac{e}{m_{\omega}}g_{\omega\sigma\gamma}\partial^\alpha
\omega^\beta(\partial_\alpha A_\beta-\partial_\beta A_\alpha
)\sigma
\end{equation}
describing the $\omega\sigma\gamma$-vertex \cite{R12}. The
electronic decay width of $\omega$ meson, neglecting the electron
mass is given as $\Gamma(\omega\rightarrow
e^+e^-)=(4\pi\alpha^2/3m_\omega^3)\lambda_\omega^2$, using the
experimental value of the decay width $\Gamma(\omega\rightarrow
e^+e^-)=(0.60\pm 0.02)$ KeV of $\omega$ meson \cite{R6}, we
determine the overlap amplitude $\lambda_\omega$ of $\omega$ meson
as $\lambda_\omega=(0.036\pm 0.001)~~GeV^2$. In a previous work,
we estimated the overlap amplitude $\lambda_\sigma$ employing a
QCD sum rule analysis of the scalar current by considering the
two-point scalar current correlation function  as
$\lambda_\sigma=(0.12\pm 0.03)~~GeV^2$ \cite{R13} since this
amplitude is not available experimentally. The
$<\sigma\gamma|\omega>$ matrix element, using Eq. 3, can be
written as $<\sigma(p^\prime)\gamma(q)|\omega(p)>=
i\frac{e}{m_\omega}g_{\omega\sigma \gamma}K(q^2)(p\cdot q
~u\cdot\epsilon -u\cdot q~ p\cdot\epsilon)$ where $q=p-p^\prime$,
$\epsilon_\mu$ is the polarization vector of the photon, and
$K(q^2)$ is a form factor with K(0)=1.

The theoretical part of the sum rule for the coupling constant
g$_{\omega\sigma\gamma}$ is obtained in terms of QCD degrees of
freedom by calculating the two point correlator in the deep
Euclidean region where ${p^\prime}^2$ and $(p^\prime+q)^2$ are
large and negative. In this calculation the full light propagator
with both perturbative and nonperturbative contributions is used,
and it is given as \cite{R14}
\begin{eqnarray}\label{e4}
  iS(x,0)&=&<0|T\{\overline{q}(x)q(0)\}|0>\nonumber \\
         &=&i\frac{\not x}{2\pi^2x^4}-\frac{<\overline{q}q>}{12}-
         \frac{x^2}{192}m_0^2<\overline{q}q>\nonumber \\
         &~&-ig_s\frac{1}{16\pi^2}\int_0^1du\left\{
         \frac{\not{x}}{x^2}\sigma_{\mu\nu}G^{\mu\nu}(ux)
         -4iu\frac{x_\mu}{x^2}G^{\mu\nu}(ux)\gamma_\nu\right\}+...
\end{eqnarray}
where terms proportional to light quark mass m$_u$ or m$_d$ are
neglected. After a straightforward computation we obtain
 \begin{eqnarray}\label{e5}
T_{\mu}(p,q)=2i\int d^4xe^{ipx}A(x_\sigma g_{\mu\tau}-x_\tau
g_{\mu\sigma})<\gamma(q)\mid
\overline{q}(x)\sigma_{\tau\sigma}q(0)\mid 0>
\end{eqnarray}
where $A=i/(2\pi^2x^4)$, and higher twist corrections are
neglected since they are known to make a small contribution
\cite{R5}. In order to evaluate the two point correlation function
further, we need the matrix elements
$<\gamma(q)|\overline{q}\sigma_{\alpha\beta}q|0>$. These matrix
elements are defined in terms of the photon wave functions
\cite{R15,R16,R17}
\begin{eqnarray}\label{e6}
<\gamma(q)|\overline{q}\sigma_{\alpha\beta}q|0>
&=&ie_q<\overline{q}q> \int_0^1due^{iuqx}\{(\epsilon_\alpha
q_\beta-\epsilon_\beta q_\alpha)\left[\chi\phi(u)+x^2[
g_1(u)-g_2(u)]\right]  \nonumber\\ && + \left[q\cdot
x(\epsilon_\alpha x_\beta-\epsilon_\beta x_\alpha)+\epsilon\cdot
x(x_\alpha q_\beta-x_\beta q_\alpha)\right]g_2(u)\}~~,
\end{eqnarray}
where the parameter $\chi$ is the magnetic susceptibility of the
quark condensate and $e_q$ is the quark charge, $\phi(u)$ stand
for the leading twist-2 photon wave function, while $g_1(u)$ and
$g_2(u)$ are the two-particle photon wave functions of twist-4. In
the further analysis the path ordered gauge factor is omitted
since we work in the fixed point gauge \cite{R18}.

The two point function T$_\mu(p,q)$ satisfies a dispersion
relation and we saturate this dispersion relation by inserting a
complete set of one hadron states into the correlation function.
This way we construct the phenomenological part of the two point
correlation function as
\begin{equation}\label{e7}
  T_{\mu}(p,q)=\frac{<\sigma\gamma|\omega>
  <\omega|j_\mu^\omega|0><0|j_{\sigma}|\sigma>}
   {(p^2-m^2_\omega)({p^\prime}^2-m^2_\sigma)}+...
\end{equation}
where the contributions from the higher states and the continuum
starting from some threshold $s_0$ are denoted by dots. In order
to take these contributions into account we invoke the
quark-hadron duality prescription and replace the hadron spectral
density with the spectral density calculated in QCD. In accordance
with the QCD sum rules method strategy, we then equate the two
representations of the two point correlation function, theoretical
and phenomenological, and construct the corresponding sum rule for
the coupling constant g$_{\omega\sigma\gamma}$.

After evaluating the Fourier transform and then performing the
double Borel transformation with respect to the variables
$Q_1^2=-{p^\prime}^2$ and $Q_2^2=-(p^\prime+q)^2$, we finally
obtain the following sum rule for the coupling constant
g$_{\omega\sigma\gamma}$
\begin{eqnarray}\label{e8}
g_{\omega\sigma\gamma}= \frac{1}{6}\frac{m_\omega
(e_u+e_d)<\overline{u}u>}{\lambda_\omega\lambda_\sigma}
  e^{m_\sigma^2/M_1^2}e^{m_\omega^2/M_2^2}
 \left\{-M^2\chi\phi(u_0)f_0(s_0/M^2)+4g_1(u_0)\right\}
\end{eqnarray}
where the function $f_0(s_0/M^2)=1-e^{-s_0/M^2}$ is the factor
used to subtract the continuum, $s_0$ being the continuum
threshold, and
\begin{equation}\label{e9}
  u_0=\frac{M_2^2}{M_1^2+M_2^2}~~~~~~,
  M^2=\frac{M_1^2M_2^2}{M_1^2+M_2^2}\nonumber
\end{equation}
with M$_1^2$ and M$_2^2$ are the Borel parameters.

The various parameters we adopt for the numerical evaluation of
the sum rule for the coupling constant g$_{\omega\sigma\gamma}$
are $<\overline{u}u>=(-0.014\pm 0.002)~~GeV^3$ \cite{R19} for the
vacuum condensate, and $\chi=-4.4~~GeV^{-2}$ \cite{R16,R20} for
the magnetic susceptibility of the quark condensate,
$\lambda_\omega=(0.036\pm 0.001)~~GeV^2$ as determined from the
experimental electronic decay width of $\omega$ meson as discussed
above, $\lambda_\sigma=(0.12\pm 0.03)~~GeV^2$ which was determined
using QCD sum rules method  \cite{R13}, and
$m_{\omega}=0.782~~GeV$, $m_{\sigma}=0.5~~GeV$. The leading
twist-2 photon wave function is given as $\phi(u)=6u(1-u)$ and the
two-particle photon wave function of twist-4 is given by the
expression $g_1(u)=-(1/8)(1-u)(3-u)$ \cite{R16}. We then study the
dependence of the sum rule for the coupling constant
g$_{\omega\sigma\gamma}$ on the continuum threshold $s_0$ and on
the Borel parameters $M_1^2$ and $M_2^2$ by considering
independent variations of these parameters. We find that the sum
rule is quite stable for $M_1^2=1.2~~GeV^2$ and for $1.0
~~GeV^2<M_2^2<1.4 ~~GeV^2$ where these limits on $M_2^2$ are
within the allowed interval for the vector channel \cite{R21}. The
variation of the coupling constant as a function of the Borel
parameter $M_2^2$ for the values of $s_0=1.1, 1.2, 1.3~~GeV^2$
with $M_1^2=1.2~~GeV^2$ is shown in Fig. 1. The sources
contributing to the uncertainty in the estimated value of the
coupling constant are those due to variations in the Borel
parameters $M_1^2$ and $M_2^2$, in the threshold parameter $s_0$,
in the estimated values of the vacuum condensate and the magnetic
susceptibility of the quark condensate, and in the values of the
overlap amplitudes $\lambda_\sigma$ amd $\lambda_\omega$. We take
these uncertainties  by a conservative estimate into account, we
obtain the coupling constant g$_{\omega\sigma\gamma}$ as $\mid
g_{\omega\sigma\gamma}\mid=(0.72\pm 0.08)$.

\begin{figure}[b]
\vspace*{-5.0cm} \hspace*{0.0cm}
\epsfig{figure=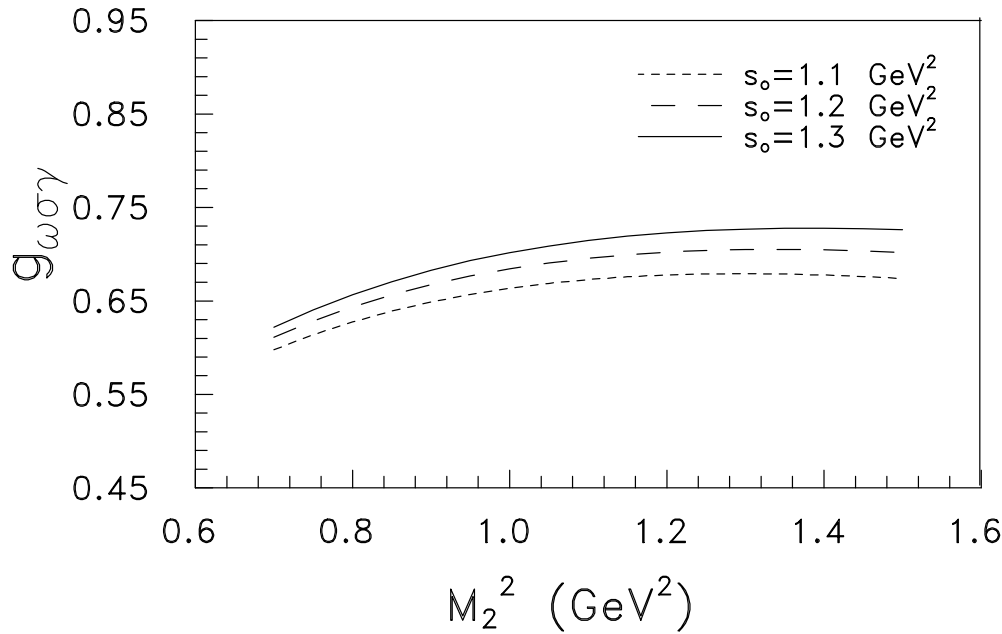,height=20cm} \vspace*{-5.0cm} \caption{The
coupling constant $g_{\omega\sigma \gamma}$ as a function of the
Borel parameter $M_2^2$ for different values of the threshold
parameter $s_0$ with $M_1^2$=1.2 GeV$^2$.}
\end{figure}

In a previous work \cite{R22}, we studied the
$\omega\rightarrow\pi\pi\gamma$ decays by adding the amplitude of
$\sigma$-meson intermediate state to the amplitude calculated
within the framework of chiral perturbation theory and vector
meson dominance. We used the experimental value for the decay rate
$\Gamma(\omega\rightarrow\pi^0\pi^0\gamma)$ and we calculated the
coupling constant  g$_{\omega\sigma\gamma}$ as a function of the
$\sigma$ meson parameters $m_\sigma$ and $\Gamma_\sigma$. If we
use the values for these parameters the values $m_\sigma=478~~MeV$
and $\Gamma_\sigma=324~~MeV$ as suggested by the Fermilab E791
Collaboration \cite{R7}, we then obtain the coupling constant
g$_{\omega\sigma\gamma}$ in the framework of this phenomenological
analysis as g$_{\omega\sigma\gamma}=0.13$ and
g$_{\omega\sigma\gamma}=-0.27$, since the theoretical calculation
of the decay rate results in a quadratic expression for the
coupling constant g$_{\omega\sigma\gamma}$. We note that these
values are consistent with the interval of values for this
coupling constant that we deduced from the experimental upper
limit of the $\Gamma(\omega\rightarrow\pi^+\pi^-\gamma)$ decay
rate which is $1.20>g_{\omega\sigma\gamma}>-1.34$ for the above
values of the $\sigma$ meson parameters \cite{R22}. The result of
our present calculation utilizing QCD sum rules method is in
reasonable agreement with the results obtained from the
phenomenological analysis of the $\omega\rightarrow\pi\pi\gamma$
decays. However, we note that in the present work we consider the
$\sigma$ meson in the narrow resonance limit and do not take its
finite width into account. Since $\sigma$ meson has a large width
\cite{R7}, the use of simple pole spectrum is somewhat doubtful.
The effect of the width of the resonance can be taken into account
within the framework of traditional QCD Laplace sum rules method,
but in the light cone QCD sum rules which forms the appropriate
approach for the present problem it seems that the Laplace
transform method cannot be used. Therefore, in our analysis the
error that is induced by the variation of threshold, which is
included in the final quoted error above, may be considered as an
estimation of the error resulting from the narrow width
approximation.

\end{document}